\documentclass[reprint, nobibnotes, amsmath, amssymb, aps, pra, floatfix, superscriptaddress]{revtex4-2}
\pdfoutput=1
\usepackage{graphicx}
\usepackage{dcolumn}
\usepackage{bm}
\usepackage[hidelinks]{hyperref}
\begin{document}

\title{Dead-zone-free single-beam atomic magnetometer based on free-induction-decay of Rb atoms}

\author{Shrey Mehta}
\affiliation{Division of Mathematical and Physical Sciences, School of Arts and Sciences, Ahmedabad University,  Ahmedabad, Gujarat 380009, India}

\author{G. K. Samanta}
\affiliation{Photonic Sciences Lab., Physical Research Laboratory, Ahmedabad 380009, Gujarat, India}

\author{Raghwinder Singh Grewal}
    \email{raghwinder.singh@ahduni.edu.in}
    
\affiliation{Division of Mathematical and Physical Sciences, School of Arts and Sciences, Ahmedabad University,  Ahmedabad, Gujarat 380009, India}

\begin{abstract}
Free-induction-decay (FID) magnetometers have evolved as simple magnetic sensors for sensitive detection of unknown magnetic fields. However, these magnetometers suffer from a fundamental problem known as a "dead zone," making them insensitive to certain magnetic field directions. Here, we demonstrate a simple experimental scheme for the dead-zone-free operation of a FID atomic magnetometer. Using a single laser beam containing equal strength of linear- and circular-polarization components and amplitude-modulation at a low-duty cycle, we have synchronously pumped the rubidium-87 atoms with both first- and second-order frequency harmonics. Such a pumping scheme has enabled us to observe the free Larmor precession of atomic spins at a frequency of $\Omega_L$ (orientation) and/or 2$\Omega_L$ (alignment) in a single FID signal, depending on the direction of the external magnetic field. We observed that the amplitude of the FID signal does not go to zero for any magnetic field direction, proving the absence of dead zones in the magnetometer. The magnetometer has a sensitivity in the range of 3.2 - 8.4 pT/$\sqrt{Hz}$ in all directions. Our experimental scheme can be crucial in developing miniaturized atomic magnetometers for various practical applications, including geomagnetic applications.
\end{abstract}

\maketitle 

Atomic magnetometers are highly sensitive devices capable of detecting very weak magnetic fields with a sensitivity level up to sub-femtotesla \cite{budker_sensitive_2000}. They operate by measuring the Larmor precession of spin-polarized atoms through a magnetic resonance signal in the presence of an external magnetic field. Atomic magnetometers have applications in a broad range of fields, including biomedical imaging \cite{ xia_magnetoencephalography_2006}, geomagnetic field mapping \cite{pedreros_bustos_remote_2018}, and space-based scientific missions \cite{korth_miniature_2016}. So far, different types of atomic magnetometers have been developed using various methods, including coherent population trapping (CPT) \cite{stahler_picotesla_2001}, nonlinear magneto-optical rotation (NMOR) \cite{budker_sensitive_2000}, free-induction decay (FID) \cite{hunter_free-induction-decay_2018, grujic_sensitive_2015}, and the Bell-Bloom configuration \cite{bloom_principles_1962}. However, all these atomic magnetometers lose sensitivity to the external magnetic field in certain directions, referred to as the "dead zones". 

These dead zones are inherent in the atomic magnetometers as the generation process of atomic-spin polarization (i.e., atomic coherence) in atoms is vectorial by nature. As a result, depending on the light polarization, atomic coherence $|\Delta m|$  of different orders can be generated by coupling two or more Zeeman sublevels \cite{ grewal_light-ellipticity_2020, gal_zero-field_2022, grewal_effect_2023}. For example, the coherence $|\Delta m|$ = 1 can be created by pumping atoms with a circularly polarized light, resulting in a magnetic resonance signal at Larmor frequency $\Omega_L$ (orientation) for the external magnetic field perpendicular to the light propagation direction. Similarly, the coherence $|\Delta m|$ = 2, produced by linearly polarized light, can generate a magnetic resonance at frequency 2$\Omega_L$ (alignment) when the external magnetic field is parallel to the direction of light propagation. Although both resonances ($\Omega_L$ \& 2$\Omega_L$) are commonly used in the atomic magnetometry \cite{fabricant_how_2023}, but their amplitude vanishes for certain magnetic field directions, resulting in dead zones.

Over the years, different techniques have been proposed and implemented to eliminate the dead zones of the atomic magnetometers designed based on various schemes \cite{ben-kish_dead-zone-free_2010, bevilacqua_magneto-optic_2014, wang_magneto-optical_2020, wu_dead-zone_2015, wang_dual-mode_2021, yu_light-shift-free_2022, tian_dead-zone_2024, tian_dead-zone-free_2024}. These techniques include polarization modulation of a laser beam \cite{ben-kish_dead-zone-free_2010, bevilacqua_magneto-optic_2014, wang_magneto-optical_2020}, use of a liquid crystal polarization rotator to keep the light polarization perpendicular to the external field \cite{wu_dead-zone_2015}, and employing an unmodulated hybrid Poincare beam to produce magnetic resonances \cite{tian_dead-zone-free_2024}. However, the elimination of dead zones in FID atomic magnetometers has not been previously studied. Therefore, it is very important to explore different techniques to develop a FID magnetometer without dead zones.

\begin{figure}[hbt!]
\centering\includegraphics[scale=0.19]{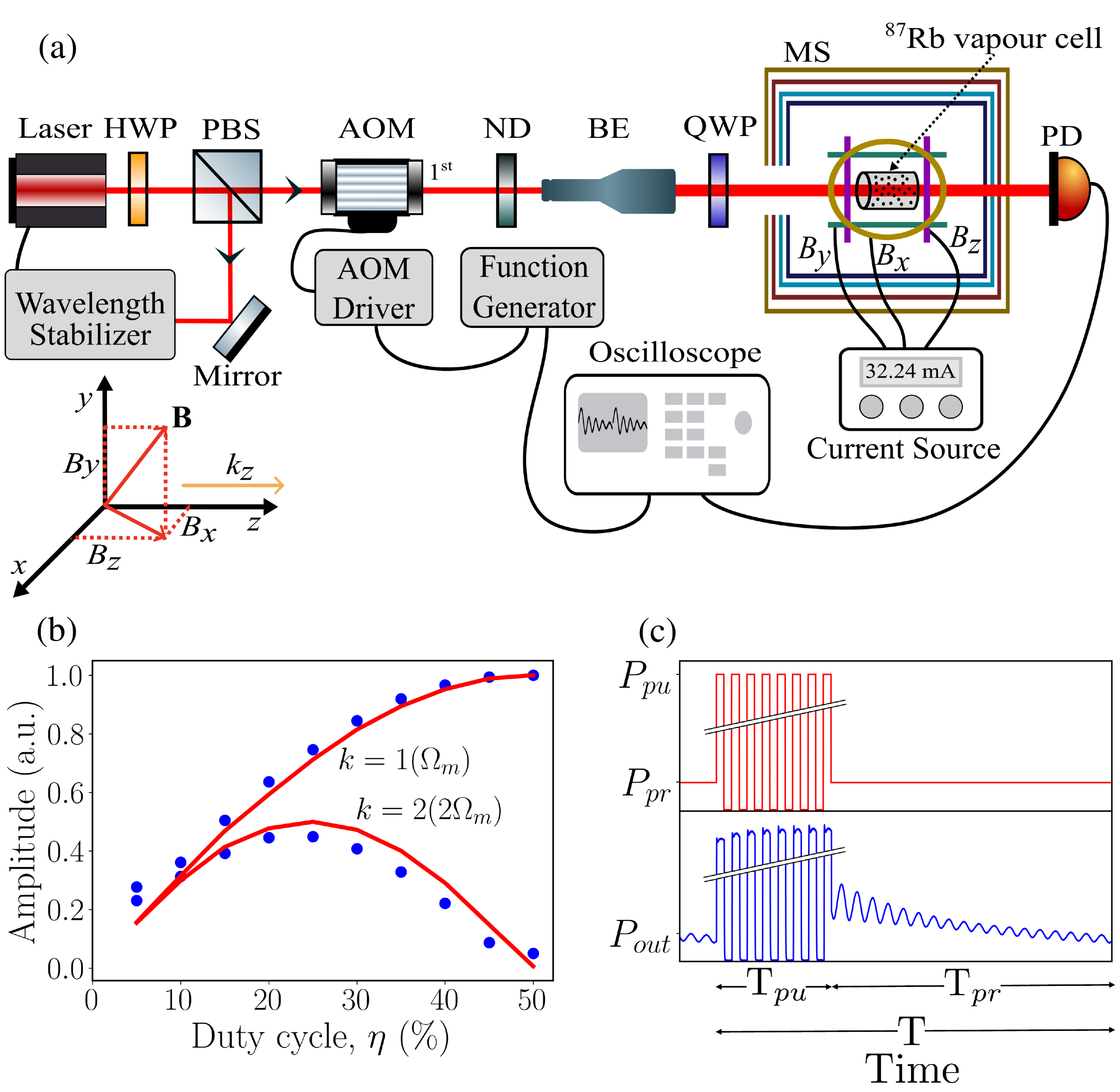}
\caption{(a) Schematic diagram of the experimental setup. HWP, half-wave plate; PBS, polarizing beam splitter; AOM, acousto-optic modulator; BE, beam expander; QWP, quarter-wave plate; ND, neutral density filter; PD, photodiode; MS, magnetic shield. (b) Normalized amplitude of first (\textit{k} = 1) and second (\textit{k} = 2) harmonics in the intensity-modulated beam as a function of duty cycle $\eta$. (c) Synchronous optical pumping configuration with a single pump-probe cycle with a measurement time T (= T$_{pu}$ + T$_{pr}$) for each FID signal (not to scale). 
}
\label{expsetup}
\end{figure}

In this work, we propose a simple technique to remove the dead zones of a FID atomic magnetometer. We have used a single laser beam modulated at different duty cycles for synchronous optical pumping of atoms and measured the FID signals by detecting the transmitted light through a rubidium ($^{87}$Rb) vapor cell filled with neon buffer gas. Subsequently, by transforming the polarization of the input light into elliptical polarization, we show that the FID magnetometer is free from any dead zones for the duty cycle of the light modulation lowered to 20$\%$. Finally, we measured the sensitivity of the dead-zone-free magnetometer as low as 3.2 pT/$\sqrt{Hz}$ for the external field applied along the light propagation direction and similar sensitivity for the external field in other directions.

The schematic of the experimental setup for the dead-zone-free FID magnetometer is shown in Fig. \ref{expsetup}(a). A diode laser (DL pro, Toptica) at a wavelength of 795 nm is used as the primary source for the experiment. The propagation direction $k_z$ of the laser beam is considered along the \textit{z}-axis. Using a combination of a half-wave plate (HWP) and a polarizing beam splitter (PBS), we controlled the laser intensity in the experiment and used the reflected beam of the PBS in combination with the wavelength stabilizer comprised of a saturation absorption spectroscopy (SAS) setup, and a PID controller to lock the laser frequency to $F_{g}$=2 $\rightarrow$$F_e$=1 transition of  $^{87}$Rb D$_1$ line.

The laser beam to the experiment is modulated at a frequency $\Omega_m$ using an acoustic-optic modulator (AOM) driven by a rectangular pulse of different duty cycles, $\eta$. The intensity of the modulated laser beam is further controlled using a neutral density (ND) filter and, subsequently, expanded to a beam diameter of 4 mm with the help of a beam expander (BE) to increase the volume of the light atom interaction region. A quarter-wave plate (QWP) is used to adjust the polarization (i.e., ellipticity $\epsilon$) of light passing through a cylindrical vapor cell containing $^{87}$Rb atoms with 10 Torr of Neon buffer gas. The vapor cell, maintained at a constant temperature of 40 $^\circ$C using a heater oven, is placed inside a four-layered $\mu$-metal magnetic shield (MS) to reduce the ambient magnetic field. Further, a three-axis printed coil driven by a multichannel current source is installed inside the $\mu$-metal shield to cancel any remaining stray field and also to apply any external magnetic field $\textbf{\textit{B}}$ ($B_x,B_y,B_z$). While the magnitude of the field $|\textbf{\textit{B}}|$ is kept constant for all measurements, the direction of the magnetic field is changed by adjusting the current in the coils. The presence of the external field $\textbf{\textit{B}}$ shifts the energy of the Zeeman sublevels $m_{F}$ by $\pm m_{F} \Omega_{{L}}$, where the term $\Omega_{{L}}= \gamma |\textbf{\textit{B}}|$ is the Larmor frequency corresponding to field  $\textbf{\textit{B}}$ and the $\gamma$ (= 0.7 MHz/G) represents the gyromagnetic ratio of the Rb atom. A photo-diode (PD), in combination with the digital oscilloscope, is used to detect the transmitted light and store the FID signal,  $\rm{S}_{\textit{FID}}$ (\textit{t}) as a function of time.

Next, we characterized the light spectrum while modulating at different duty cycles. According to the Fourier series, the change in the duty cycle, $\eta$, of a periodic function of frequency, $\Omega_m$, results in a change in the amplitude of the harmonics of frequency, $k \Omega_m$, where the integer $k$ corresponds to the order of the sideband or harmonics of the pulse \cite{ grewal_light-ellipticity_2020}. Therefore, keeping the modulation frequency, $\Omega_m$, fixed, we varied the duty cycle, $\eta$, and measured the amplitude of the first-harmonic ($\Omega_m$) and second-harmonic ($2\Omega_m$) of the modulation frequency components in the beam. The results are shown in Fig. \ref{expsetup}(b). As expected, for $\eta$ = 50$\%$, the normalized amplitude of the first and second harmonic are maximum and minimum, respectively. The amplitude (blue dots) of the first and second harmonic decrease and increase, respectively, with the decrease of $\eta$ with a comparable amplitude at around $\eta$ = 20$\%$ in close agreement with the theoretical values (line). A further decrease in $\eta$ reduces both harmonics due to the transfer of energy to higher harmonics.

To confirm the generation of FID in the current experiment, we have modulated the input laser beam using the AOM driven by the square wave signal of $\eta$ = 50$\%$ and recorded the PD signal after the vapor cell. The results are shown in Fig. \ref{expsetup}(c). The generation of a FID signal is a two-step process \cite{grujic_sensitive_2015,hunter_free-induction-decay_2018}: the pump phase and the probe phase [see red curve of Fig. \ref{expsetup}(c)]. During the pump phase (i.e., for a time duration, T$_{pu}$),  the atoms are synchronously pumped by a polarized light whose amplitude is modulated at the Larmor frequency $\Omega_{{L}}$ or 2$\Omega_{{L}}$ in the presence of field $\textbf{\textit{B}}$. The synchronous optical pumping is performed at peak pump power $P_{pu}$. This creates a long-lived Zeeman coherence among the hyperfine ground-state sublevels, resulting in a spin-polarized medium. During the probe phase, a time duration, T$_{pr}$, light modulation is turned off, and the laser power is reduced to a constant level $P_{pr}$ (see red curve of Fig. \ref{expsetup}(c)). This allows the free Larmor precession of atomic spins, which can be observed as a damped oscillation (blue curve of Fig. \ref{expsetup}(c)) in the transmitted light $P_{out}$, i.e., a FID signal, in time T$_{pr}$.

The atomic spins can precess at Larmor frequency $\Omega_{{L}}$ (orientation) and/or 2$\Omega_{{L}}$ (alignment) depending on the relative angle between the polarization vector and the magnetic field vector. The behavior of a single FID signal can be described by \cite{lenci_vectorial_2014, grewal_effect_2023}
\begin{eqnarray}
S_{FID}(t) =b_o + b_{s}  e^{-\gamma_s t} +  e^{-\gamma_o t}[b_{\Omega_L} \mathrm{sin}(2 \pi \Omega_L t + \phi) \nonumber \\ + b_{2\Omega_L} \mathrm{sin}(4 \pi \Omega_L t + \phi)] 
\label{fittingeq}
\end{eqnarray}

\noindent where $\gamma_s$ and $\gamma_o$ are the decay rates of static and oscillatory parts of the FID signal, respectively. The term $b_0$ is the time-independent transmission in the signal. The terms $b_s$, $b_{\Omega_L}$, and $b_{2\Omega_L}$ are the amplitudes of corresponding static, $\Omega_{{L}}$ and 2$\Omega_{{L}}$ oscillations, respectively. The constant $\phi$ is the initial phase of both oscillations \cite{grewal_transient_nmor2020}. Further, we fitted Eq. \ref{fittingeq} to the experimentally recorded FID signal measured at different directions of the magnetic field, and extracted the fitting parameters, including the amplitudes $b_{\Omega_L}$ and $b_{2\Omega_L}$, and the Larmor frequency $\Omega_{{L}}$ corresponding to the external magnetic field, B.

\begin{figure}[hbt!]
    \centering
    \includegraphics[width=0.93\linewidth]{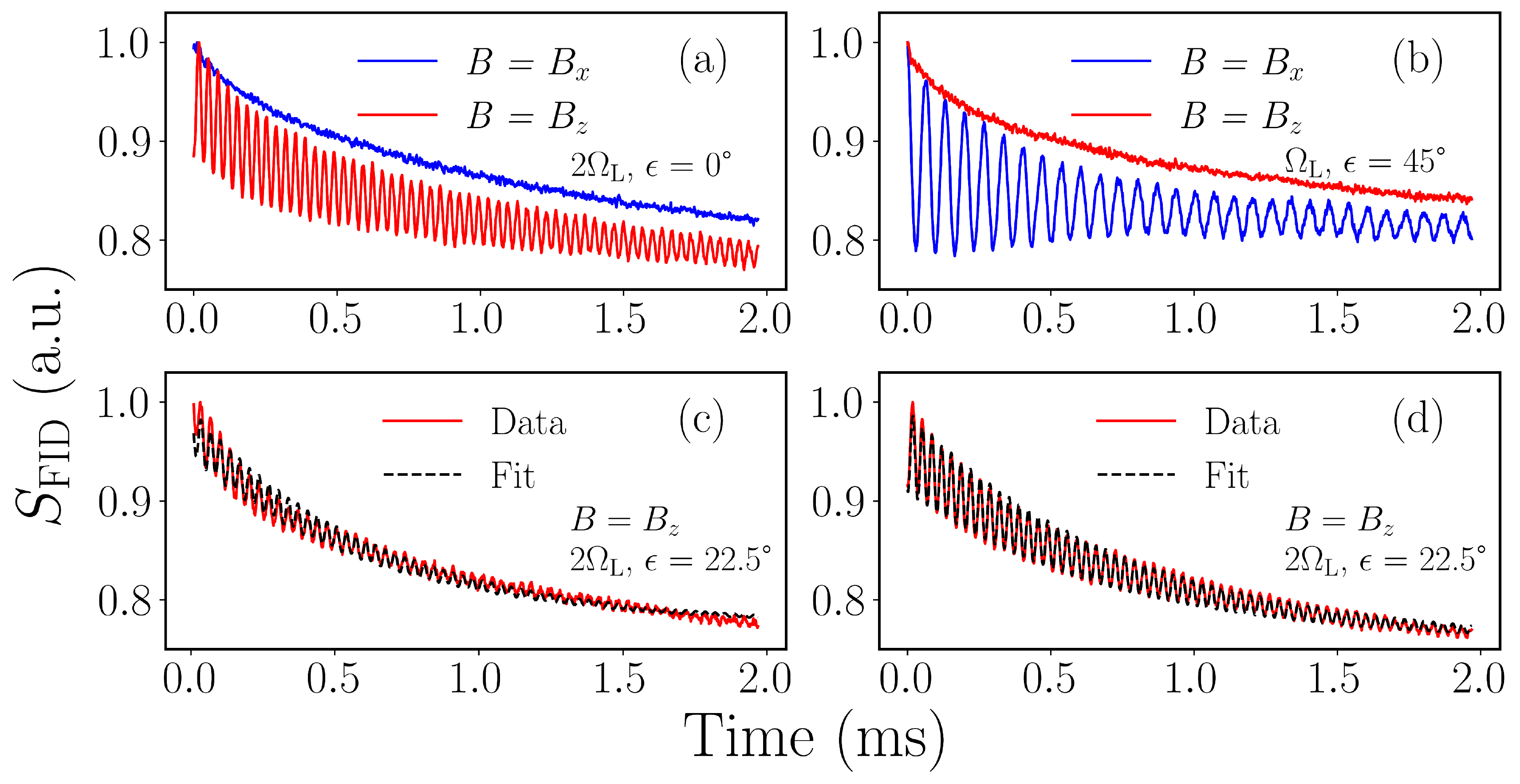}
    \caption{FID signals measured at different experimental conditions. Duty cycle, $\eta$ of light modulation, (a) to (c) 50$\%$ and (d) 20$\%$. (c) and (d) The dashed black line shows the fit to the FID signals.}
    \label{FIDPlots}
\end{figure}

First, we used linearly and circularly polarized light beams modulated at duty cycle $\eta$= 50$\%$ for optical pumping in the FID magnetometer. Applying a constant magnetic field $\textbf{\textit{B}}$ = 21.43 mG (i.e., $\Omega_{{L}}$ = 15 kHz) along \textit{z}-axis (i.e., $\textbf{B}\parallel k_z$), we modulated the light intensity at a frequency, $\Omega_m$ = 30 kHz to match with the oscillation frequency 2$\Omega_{{L}}$ (= 30 kHz)(alignment) of the FID signal. As evident from Fig. \ref{FIDPlots}(a), the FID signal (red curve) recorded for the linear polarized light ($\epsilon$ = 0$^\circ$) with polarization vector aligned along the \textit{x}-axis has a characteristic oscillation confirming the sensitivity of the FID magnetometer for \textit{B} = $B_z$. However, as we align the field $\textbf{\textit{B}}$ along the direction of the light polarization vector, i.e., \textit{B} = $B_x$, we do not see any oscillation (blue curve) in the signal. Such observation confirms the presence of a "dead zone" for the FID atomic magnetometer pumped with linearly polarized light. On the other hand, as evident from Fig. \ref{FIDPlots}(b), we can observe the FID signal for \textit{B} = $B_x$ (blue curve) when pumped by circularly polarized light ($\epsilon$ = 45$^\circ$). In this case, the FID signal oscillates at $\Omega_{{L}}$ (orientation), so we fix the light modulation frequency $\Omega_m$ = 15 kHz for synchronous optical pumping (i.e., $\Omega_m$ = $\Omega_{{L}}$) \cite{grujic_sensitive_2015, hunter_free-induction-decay_2018}. However, the FID signal has no characteristic oscillation when the atomic medium is pumped by circularly polarized light, and the magnetic field is along propagation of light, \textit{B} = $B_z$ (red curve), confirming the dead zone of the FID magnetometer.

From Fig. \ref{FIDPlots}(a) and \ref{FIDPlots}(b), it is evident that the FID magnetometers have dead zones for both linearly ($\epsilon$ = 0$^\circ$) and circularly ($\epsilon$ = 45$^\circ$) polarized pump light beams, but for different directions of $\textbf{\textit{B}}$. These observations also indicate the possibility of removal of such dead zones by simultaneously pumping the atoms by circular and linear polarized light. However, the FID signal oscillates at Larmor frequencies $\Omega_{{L}}$ and/or 2$\Omega_{{L}}$ depending on the field $\textbf{\textit{B}}$ direction. As a result, to achieve the synchronous optical pumping for $\Omega_{{L}}$ and 2$\Omega_{{L}}$ oscillations, the input beam modulated at fixed frequency $\Omega_m$ must contain both first and second harmonic of comparable amplitudes to satisfy $\Omega_m$ = $\Omega_{{L}}$ and 2$\Omega_m$ = 2$\Omega_{{L}}$. Our experimental setup (see Fig. \ref{expsetup}(a)) addresses all these requirements with ease. Using the QWP placed before the gas cell, we set the light ellipticity $\epsilon$ to 22.5$^\circ$ ensuring the input beam carries an equal amplitude of linear- and circular-polarization components. We further adjusted the amplitude of $\Omega_m$ ($k = 1$) and 2$\Omega_m$ ($k = 2$) harmonic components in the modulated pump beam by simply controlling its duty cycle $\eta$ (see Fig. \ref{expsetup}(b)). 

To test our proposed method, we fixed the light ellipticity $\epsilon$ = 22.5$^\circ$ and measured the FID signals by applying the external applied $\textbf{\textit{B}}$ (i.e., $\Omega_{{L}}$ = 15 kHz) along the \textit{z}-axis for two different duty cycles, $\eta$ = 50$\%$ and $\eta$ = 20$\%$, with the results shown in Fig. \ref{FIDPlots}(c) and  Fig. \ref{FIDPlots}(d), respectively. The light intensity is modulated at frequency $\Omega_m$ = 15 kHz. We kept the peak pump power $P_{pu} \approx$ 574 $\mu$W and probe power $P_{pr} \approx$ 10 $\mu$W. As evident from Fig. \ref{FIDPlots}(b), for $B = B_z$, and $\eta$ = 50$\%$, the circular polarization ($\epsilon$ = 22.5$^\circ$) component of light does not contribute to the FID signal due to presence of dead zone along \textit{z}-axis. Therefore, the FID signal (red curve), as observed in Fig. \ref{FIDPlots}(c), arises due to the linear polarization component of the light and has an oscillation frequency of 2$\Omega_{{L}}$. However, the observed signal is weak. This is because, at 50$\%$ duty cycle, the modulated light has only the first-order sideband ($k = 1$) at frequency $\Omega_m$ (see Fig. \ref{expsetup}(b)), and the atoms can not be synchronously pumped at Larmor frequency 2$\Omega_{{L}}$ as $\Omega_m$ $\neq$ 2$\Omega_{{L}}$.

However, we observe an increase in the FID signal strength as the duty cycle $\eta$ is lowered from 50$\%$. Such effect can be understood as follows. 
As evident from Fig. \ref{expsetup}(b), the decrease of duty cycle, $\eta$, from 50$\%$ increases the amplitude of the $k = 2$ frequency sideband and thus enhancing the synchronous optical pumping at Larmor frequency 2$\Omega_{{L}}$ (= 2$\Omega_{m}$). For $\eta$ $\sim$ 20$\%$, the amplitude of the $k = 2$ frequency becomes comparable to the amplitude of $k = 1$ sideband and produces the strongest FID signal (see red curve of Fig. \ref{FIDPlots}(d)). Further decrease of $\eta$ below 20$\%$ reduces the amplitude of the $k = 2$ frequency sideband and FID signal, making $\eta$ $\sim$ 20$\%$ as the optimum duty cycle for the current study. 

\begin{figure}[hbt!]
    \centering
    \includegraphics[width=0.93\linewidth]{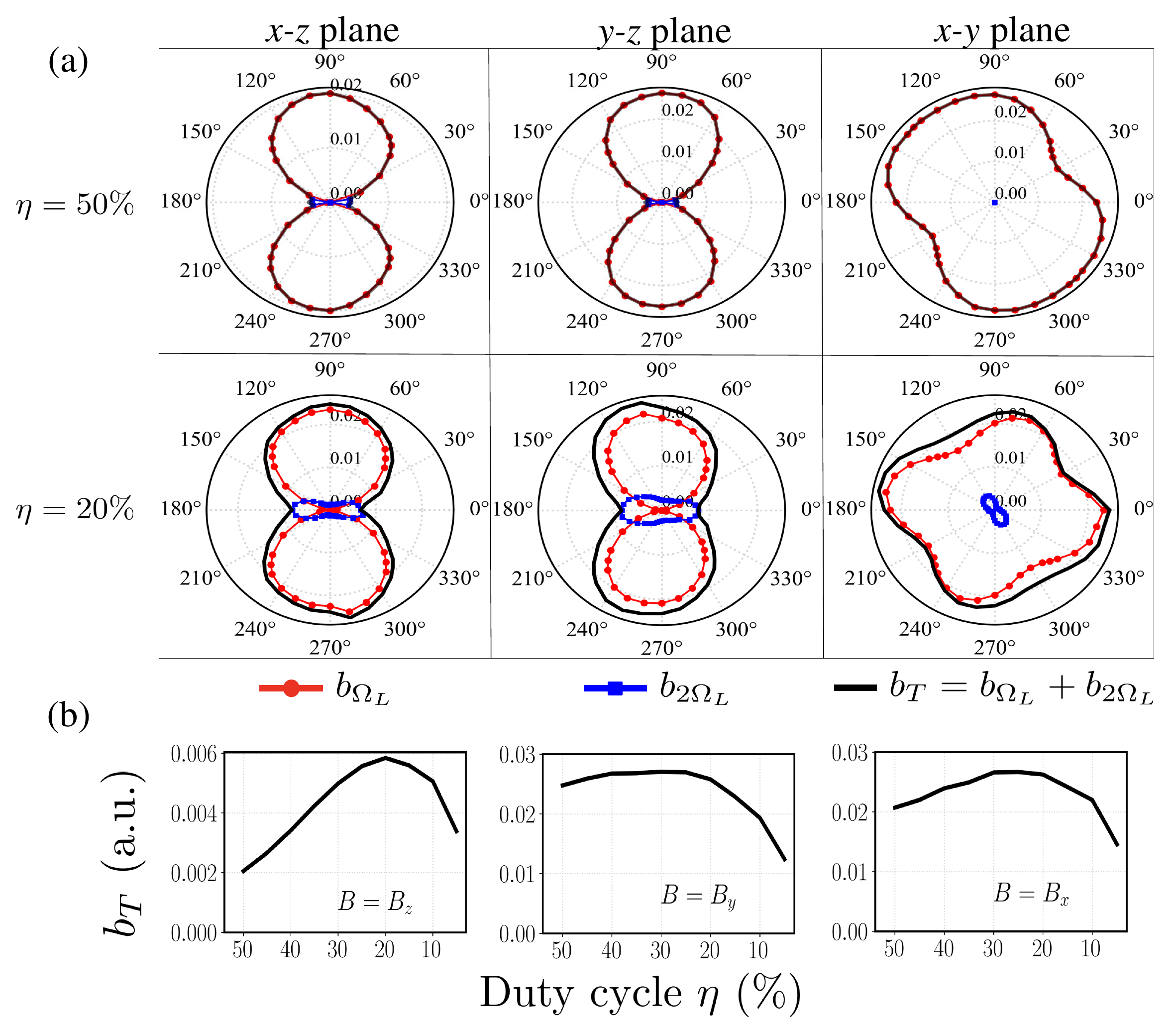}
    \caption{(a) Polar plot of the amplitudes $b_{\Omega_L}$, $b_{2\Omega_L}$  and $b_{T}$ of the FID signals with the angle, $\theta$ of the field \textbf{\textit{B}} in all three major planes for $\eta$= 50$\%$ (top row), and 20$\%$ (bottom row). (b) Variation of the total amplitude $b_{T}$ with $\eta$ for field \textbf{\textit{B}} aligned along the major axes. 
    }
    \label{polarplot}
\end{figure}

Further, we measured the amplitude of the FID signals for different orientations of the external field, $\textbf{\textit{B}}$, in three major planes (i.e., \textit{x-z}, \textit{y-z}, \textit{x-y}) at two different duty cycles, $\eta$ = 50$\%$ and $\eta$ = 20$\%$. The results are shown in Fig. \ref{polarplot}(a). We have plotted the amplitude of ${\Omega_L}$ ($b_{\Omega_L}$) and 2${\Omega_L}$ ($b_{2\Omega_L}$) oscillations along with the total amplitude $b_{T}$(= $b_{\Omega_L}$+$b_{2\Omega_L}$) of the FID signal for each plane. In  \textit{x-z} and \textit{y-z}-planes, the angle $\theta$ of field $\textbf{\textit{B}}$ is defined with respect to the \textit{z}-axis, while for \textit{x-y} plane, it is measured from the \textit{x}-axis. As evident from the top row of Fig. \ref{polarplot}(a), for $\eta$ = 50$\%$, the total amplitude $b_{T}$ (black curve) of the FID signal becomes very small when the field $\textbf{\textit{B}}$ angle is close to 0$^\circ$ or 180$^\circ$ in both \textit{x-z} and \textit{y-z} planes, i.e., $\textit{B}$ = $B_z$. This is because, for $\textit{B} = B_z$ at $\eta$ = 50$\%$, the amplitude $b_{\Omega_L}$(red curve) vanishes completely, and the amplitude $b_{2\Omega_L}$(blue curve) becomes significantly weak. So, both ${\Omega_L}$  and 2${\Omega_L}$ oscillations are not able to contribute to the FID signal simultaneously for $\textit{B}$ = $B_z$. 
 
As we tilt the magnetic field $\textbf{\textit{B}}$ either towards \textit{x}-axis or \textit{y}-axis,
the $b_{\Omega_L}$ grows significantly. This leads to an increase in the total amplitude $b_{T}$ of the FID signal along  \textit{x}- and \textit{y}-axes in all three major planes, as observed in the top row of Fig. \ref{polarplot}(a)(top row). So, the FID magnetometer pumped by elliptically polarized light modulated with a duty cycle of 50$\%$ has a single dead zone for magnetic fields making a small angle to the \textit{z}-axis (i.e., along $k_z$).

However, such dead zones can be eliminated by lowering the modulation duty cycle to 20$\%$. It is evident from the bottom row of Fig. \ref{polarplot}(a) that the total amplitude $b_{T}$ (black curve) of the FID signal does not go to zero for any magnetic field direction in all three major planes. As explained previously, lowering of $\eta$ values increases the amplitude of the $k$ = 2 (2$\Omega_m$) harmonic, enabling the synchronous optical pumping of atomic spins precessing at Larmor frequency 2${\Omega_L}$, and subsequent increase in amplitude $b_{2\Omega_L}$(blue curve) 
when the field $\textbf{\textit{B}}$ is parallel to the \textit{z}-axis. Therefore, unlike the top row of Fig. \ref{polarplot}(a), the amplitude $b_{T}$ (black curve), as shown by the bottom row of Fig. \ref{polarplot}(a), remains nonzero for all magnetic field directions.

We have further measured the variation of the total amplitude of $b_{T}$ as a function of $\eta$ for the field \textit{\textbf{B}} along three coordinate axes with the results shown in Fig. \ref{polarplot}(b). It is interesting to note that for \textit{B} = $B_z$, the amplitude $b_{T}$ increases significantly with the decrease of $\eta$ resulting in a maximum value of $b_{T}$ at $\eta$ $\sim$20$\%$. On the other hand, the $b_{T}$ grows marginally with the decrease of $\eta$ for both \textit{B} = $B_y$ and \textit{B} = $B_x$ up to $\eta$ $\sim$20$\%$. However, for $\eta$ value below 20$\%$ reduces the value of $b_{T}$ for the field \textit{\textbf{B}} along all three coordinate axes due to the decrease of the amplitude of $k$ = 1 ($\Omega_m$) and $k$ = 2 (2$\Omega_m$) harmonics (see Fig. \ref{expsetup}(b)) confirming $\eta$ $\sim$20$\%$ as the optimum duty cycle for light modulation to achieve dead-zone-free FID magnetometer.

\begin{figure}[hbt!]
    \centering
    \includegraphics[scale=0.21]{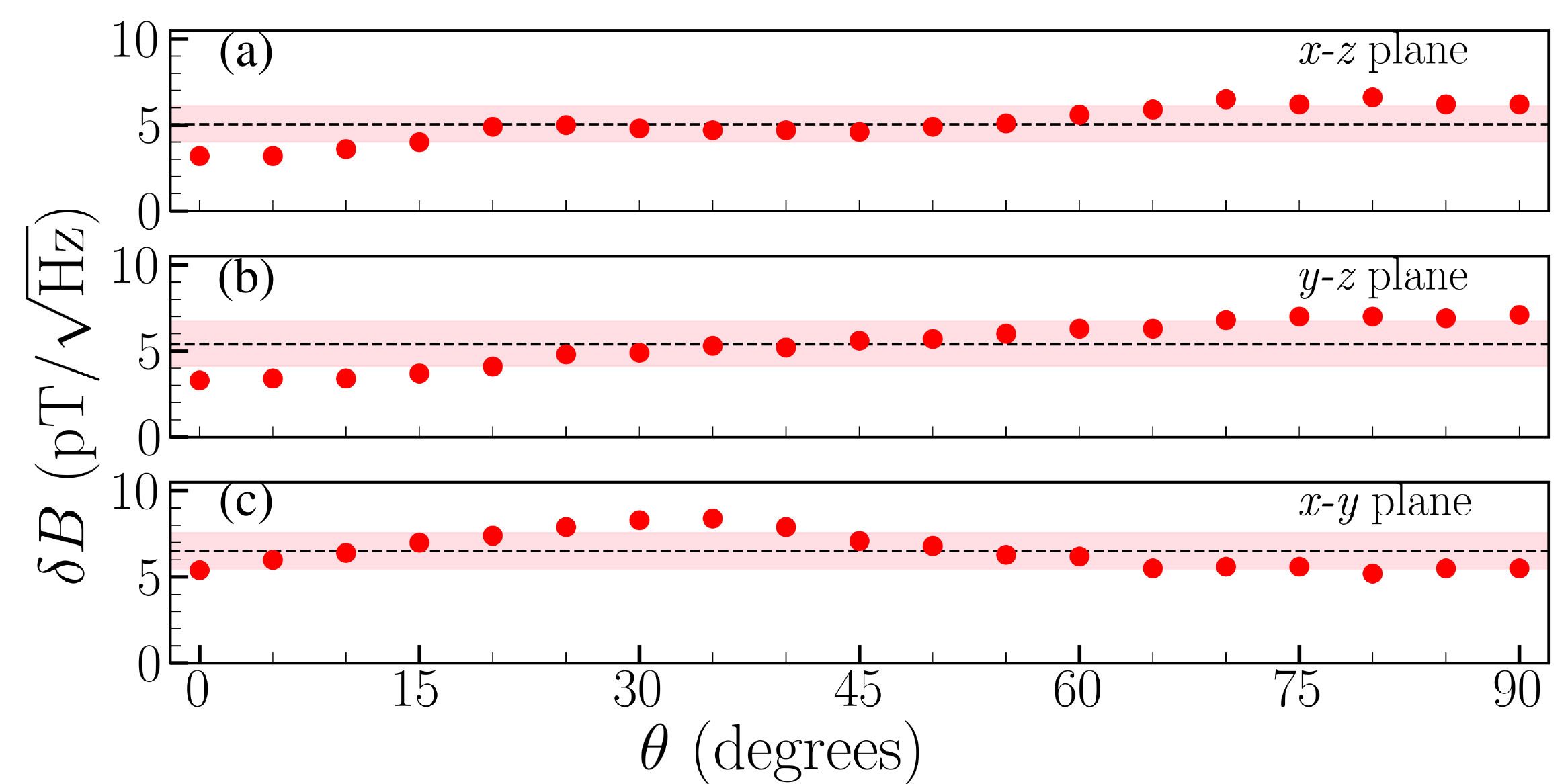}
    \caption{Variation of sensitivity $\delta B$ as a function of field \textbf{\textit{B}}  angle $\theta$ in (a) \textit{x-z}, (b) \textit{y-z}, and (c) \textit{x-y} planes. All the measurements are done at powers $P_{pu} \ \approx$  1.3 mW  and $P_{pr} \ \approx$  30 $\mu$W.
    } 
        \label{senplot}
\end{figure}

Finally, keeping $\eta$ = 20$\%$ fixed, we evaluated the sensitivity $\delta B$ of the magnetometer as a function of $\textbf{\textit{B}}$ field angle $\theta$ in all three major planes. To find the $\delta B$, we recorded 500 FID traces for a time period of one second with a measurement time of T = 2 ms for each signal. All the individual FID traces were fitted using Eq. \ref{fittingeq} to extract ${\Omega_L}$ (i.e., Larmor frequency), and the magnitude of the field \textbf{\textit{B}}. The time series data of \textbf{\textit{B}} was converted to the frequency domain by taking an FFT (Fast Fourier Transform). Finally, the magnetic-field-noise floor ($\delta B$) was calculated from the root-mean-square amplitudes of noise \cite{fabricant_how_2023} for each angle $\theta$ in all three major planes, with the results shown in Fig. \ref{senplot}

As evident from Fig. \ref{senplot}(a), the sensitivity $\delta B$ is measured as a function of angle $\theta$ varying from 0$^\circ$ to 90$^\circ$ in the \textit{x-z} plane. We obtain the $\delta B$ = 3.2 $\mathrm{pT}/\sqrt{Hz}$ at $\theta$ = 0$^\circ$ of the dead-zone-free FID magnetometer. An average sensitivity $\delta B_{avg}$ = 5.1 $\mathrm{pT}/\sqrt{Hz}$ (dashed line) is measured when the field angle $\theta$ is changed in this plane. Similarly, as shown by the dashed line of Fig. \ref{senplot}(b) and Fig. \ref{senplot}(c) corresponding to \textit{y-z} and \textit{x-y} planes, respectively, the average sensitivity of the magnetometer is almost same in all planes. The $\delta B_{avg}$ has measured values of 5.4 $\mathrm{pT}/\sqrt{Hz}$ and  6.5 $\mathrm{pT}/\sqrt{Hz}$ in \textit{y-z} and \textit{x-y} planes, respectively. It is also interesting to note that almost all the data points of $\delta B$ as shown in Fig. \ref{senplot} lie within the red-colored shaded regions representing the value 2$\times$ the standard deviation of $\delta B$ across $\delta B_{avg}$ for each plane. The overall variation in $\delta B$ is in the range of 3.2 - 8.4 pT/$\sqrt{Hz}$ in all directions. Since the sensitivity of the FID atomic magnetometer depends on various factors, including the power fluctuation noise of the laser and optimized temperature of the atomic vapor cell, one can, in principle, optimize these parameters to further enhance the sensitivity of the current FID magnetometer while taking advantage of its dead zone free performance.

In conclusion, we have demonstrated a simple dead-zone-free FID magnetometer by optimizing the intensity modulation of the laser beam with a duty cycle of 20$\%$. This approach ensures the presence of both the first ($\Omega_m$) and second harmonic (2$\Omega_m$) components in the modulated signal, with equal contributions from the linear and circular polarization of the pump beam. As a result, the atomic spins are synchronously pumped at both $\Omega_L$ and 2$\Omega_L$, producing a FID signal that oscillates at these frequencies based on the magnetic field direction. This effectively eliminates the dead zones typically found in FID magnetometers. The highest sensitivity achieved was 3.2 pT/$\sqrt{Hz}$, with consistent performance across all three major planes. Unlike other atomic magnetometers, this method does not require lock-in detection. Such a generic experimental scheme based on a single beam, along with simple control of the duty cycle and polarization, can be useful for developing miniaturized FID magnetometers for different applications.

\begin{acknowledgments}
S. M. and R. S. G. acknowledge the support of the Science and Engineering Research Board (SERB), Department of Science and Technology, Govt. of India through Grant No. SRG/2022/000322, and Ahmedabad University through Grant No. URBSASI22A4. G. K. S. acknowledges the support of the Department of Space, Govt. of India.
\end{acknowledgments}

\section*{Data Availability}
Data underlying the results presented in this paper may be obtained from the authors upon reasonable request.

\section*{References}
\bibliography{sample}

\end{document}